\documentclass[12pt]{article}
\usepackage{axodraw} 
\usepackage[dvips]{graphicx}

\def\dspace{\baselineskip = 0.30in}

\def\lapproxeq{\lower .7ex\hbox{$\;\stackrel{\textstyle
<}{\sim}\;$}}
\def\gapproxeq{\lower .7ex\hbox{$\;\stackrel{\textstyle
>}{\sim}\;$}}

\begin{document}

\dspace

\begin{titlepage}
\begin{flushright}
%\preprint{
BA-03-16\\
%March 2002\\
%}
\end{flushright}
\vskip 2cm
\begin{center}
%\title
{\Large\bf
Low Energy Consequences \\
~of Five Dimensional $SO(10)$ 
}
\vskip 1cm
{\normalsize\bf
%\author{
Bumseok Kyae\footnote{bkyae@bartol.udel.edu}, 
Chin-Aik Lee\footnote{jlca@udel.edu}, and 
Qaisar Shafi\footnote{shafi@bartol.udel.edu}
}
\vskip 0.5cm
{\it Bartol Research Institute, University of Delaware, \\Newark,
DE~~19716,~~USA\\[0.1truecm]}

%
%%\maketitle

\end{center}
\vskip .5cm

%\date{\today}
%\pacs{PACS: 11.25.Mj, 12.10.Dm, 98.80.Cq}

\begin{abstract}
We consider five dimensional (5D) supersymmetric $SO(10)$ compactified 
on the orbifold $S^1/(Z_2\times Z_2')$ such that the $SO(10)$ gauge symmetry 
is broken on both fixed points (branes), 
and the residual gauge symmetry is  
$SU(3)_c\times SU(2)_L\times U(1)_Y\times U(1)_X$. 
We explore one example in which the gauge symmetries on the two branes 
are respectively $SU(5)\times U(1)_X$ and 
$SU(4)_c\times SU(2)_L\times SU(2)_R$, 
and the MSSM gauge symmetry is recovered by the usual Higgs mechanism.  
We discuss how fermion masses and mixings can be understood 
in this framework by introducing a flavor $U(1)_F$ symmetry.  
Unification of the MSSM gauge couplings and proton stability are also 
considered. 
An order of magnitude increase in sensitivity could reveal proton decay.  

\end{abstract}
\end{titlepage}

\newpage

%%%%%%%%%
%%%%%%%%%

%%%%%%%%%%%%%%%%%%%%%%%%%%%%%%%%%%%%%%%%%%%%%%%%%%%%%%%%%%%%%%%%%%%%%%%%%%%
\section{Introduction}

Higher dimensional supersymmetric grand unified theories (SUSY GUTs) 
compactified on suitable orbifolds resolve the notorious doublet-triplet
problem and eliminate the troublesome dimension five nucleon decay process 
associated with four dimensional (4D) SUSY GUTS in a relatively painless 
manner~\cite{hn03,hn11,kawamura}.    
With the exception of just a handful of unified gauge groups 
such as $SU(4)_c\times SU(2)_L\times SU(2)_R$~\cite{pati,ps},  
$SU(3)_c\times SU(3)_L\times SU(3)_R$~\cite{su3^3} and 
(flipped) $SU(5)'\times U(1)_X'$~\cite{flipped}  
which allow rather elegant resolutions of these problems, 
most 4D SUSY GUTS require rather complicated Higgs systems and 
additional symmetries.  Thus, one or more extra dimension(s) can play 
a crucial role in the construction of realistic models based 
on gauge groups $SU(5)$~\cite{SU5}, $SO(10)$~\cite{SO10} 
and $E_6$~\cite{E6}.      

In this paper we investigate the construction and implications of 
five dimensional (5D) $SO(10)$ 
compactified on the orbifold $S^1/(Z_2\times Z_2')$ such that 
on each of the two fixed points (branes B1 and B2), 
the four dimensional gauge symmetry corresponds 
to one of the maximal subgroups of $SO(10)$~\cite{ks2}.  
Thus, after compactification,  
the residual four dimensional gauge symmetry group is
$SU(3)_c\times SU(2)_L\times U(1)_Y\times U(1)_X$.  
An additional $U(1)_X$ factor is present whose breaking is achieved 
via the standard Higgs mechanism.  
The MSSM gauge group is realized by spontaneously breaking $U(1)_X$ 
with a bulk Higgs hypermultiplet.

The plan of the paper is as follows.  
In section 2, 3 and 4 we describe the compactification scenario and 
the various symmetry breaking patterns from 5D $SO(10)$.     
Especially, in section 4 we discuss how the MSSM is realized at energies 
below $M_c$, with gauge symmetries $SU(5)\times U(1)$ and 
$SU(4)_c\times SU(2)_L\times SU(2)_R$ present 
on B1 and B2 respectively.  
Section 5 is devoted to a discussion of fermion masses and mixings, 
including the neutrino sector.  
We introduce a suitable $U(1)_F$ flavor symmetry,  
which enables us to realize the hierarchies displayed 
by the charged quark masses and mixings,
as well as bilarge mixings in the neutrino sector.  
In section 6 we discuss the gauge coupling unification and proton stability 
based on the model of section 5.  We conclude in section 7.  

\section{Orbifold Symmetry Breakings in 5D $SO(10)$}

The $SO(2n)$ generators are represented as
$\left(\begin{array}{cc}
A+C&B+S\\
B-S&A-C
\end{array}\right)$, 
where $A$,$B$, $C$ are $n\times n$ anti-symmetric matrices
and $S$ is an $n\times n$ symmetric matrix \cite{zee}.
By an unitary transformation,
the generators are given by
\begin{eqnarray} \label{general}
\left(\begin{array}{cc}
A-iS&C+iB\\
C-iB&A+iS
\end{array}\right)~, 
\end{eqnarray}
where $A$ and $S$ denote $U(n)$ generators, and
$C\pm iB$ transform under $SU(n)$ as $n(n-1)/2$ and $\overline{n(n-1)/2}$, 
respectively.   
Under $SU(5)\times U(1)_X$, the $SO(10)$ generators are decomposed as  
\begin{eqnarray} \label{so10}
T_{SO(10)}=\left[\begin{array}{c|c}
{\bf 24}_0+{\bf 1}_0& {\bf 10}_{-4}\\
\hline
{\bf \overline{10}}_{4}&{\bf \overline{24}}_0-{\bf 1}_0
\end{array}\right]_{10\times 10} ~,
\end{eqnarray}
where the subscripts labeling the $SU(5)$ representations indicate
$U(1)_X$ charges, and the subscript ``$10\times 10$'' denotes the matrix
dimension.  Also, ${\bf 24}$ ($={\bf \overline{24}}$) corresponds
to $SU(5)$ generators, while
${\rm diag}~({\bf 1}_{5\times 5},-{\bf 1}_{5\times 5})$
is the $U(1)_X$ generator.
The $5\times 5$ matrices ${\bf 24}_0$ and ${\bf 10}_{-4}$ are further
decomposed under $SU(3)_c\times SU(2)_L\times U(1)_Y$ as
\begin{eqnarray} \label{24}
&&{\bf 24}_{0}=\left(\begin{array}{cc}
{\bf (8,1)}_{0}+{\bf (1,1)}_0 & {\bf (3,\overline{2})}_{-5/6}\\
{\bf (\overline{3},2)}_{5/6} & {\bf (1,3)}_{0}-{\bf (1,1)}_0
\end{array}\right)_{0}~,  \\
&&{\bf 10}_{-4}=\left(\begin{array}{cc}
{\bf (\overline{3},1)}_{-2/3} & {\bf (3,2)}_{1/6}\\
{\bf (3,2)}_{1/6} & {\bf (1,1)}_{1}
\end{array}\right)_{-4} ~.  
\end{eqnarray}
%
%where the subscripts labelling the $SU(3)_c\times SU(2)_L$ representations
%denote $U(1)_Y$ hypercharge, with the normalized generator
%$Y_{\rm GG}\equiv {\rm diag.}(-1/3,-1/3,-1/3,+1/2,+1/2)$.
%
Thus, each representation carries two independent $U(1)$ charges.
Note that the two ${\bf (3,2)}_{1/6}$s in ${\bf 10}_{-4}$ are identified.
%
%The ${\bf 10}_{-4}$ and ${\bf \overline{10}}_4$ are antisymmetric matrices.
%${\bf \overline{24}}_0$ ($={\bf 24}_0$) and
%${\bf \overline{10}}_{4}$ are also decomposed in similar manner.
%

We intend to break $SO(10)$ to its maximal subgroups  
by $Z_2$ orbifoldings.  
Let us consider the action on $SO(10)$ of the following $Z_2$ group elements,   
\begin{eqnarray} \label{p1}
P_1&=&{\rm diag.}\bigg(+I_{3\times 3},+I_{2\times 2},
+I_{3\times 3},+I_{2\times 2}\bigg)~\longrightarrow SO(10)~, \\
P_2&=&{\rm diag.}\bigg(+I_{3\times 3},+I_{2\times 2},
-I_{3\times 3},-I_{2\times 2}\bigg)~\longrightarrow 
SU(5)\times U(1)_X, \\
P_3&=&{\rm diag.}\bigg(-I_{3\times 3},+I_{2\times 2},
+I_{3\times 3},-I_{2\times 2}\bigg)~\longrightarrow
SU(5)'\times U(1)_X'~, \\  \label{p4}
P_4&=&{\rm diag.}\bigg(+I_{3\times 3},-I_{2\times 2},
+I_{3\times 3},-I_{2\times 2}\bigg)~\longrightarrow 
SU(4)_c\times SU(2)_L\times SU(2)_R ~,   ~~~
\end{eqnarray}  
where $I$'s denote identity matrices.  
Here the $P$'s all satisfy $P^2=I_{5\times 5}$.  
Eqs.~(\ref{p1})--(\ref{p4}) show all possible ways to define 
the 10 dimensional $Z_2$ group elements and the maximal subgroups 
of $SO(10)$ obtained by their operations, as will be explained below.         

Under the operations $P_1T_{SO(10)}P_1^{-1}$, $P_2T_{SO(10)}P_2^{-1}$, 
$\cdots$, the matrix elements of $T_{SO(10)}$ transform as
\begin{eqnarray} \label{so10/z2z2}
\left[\begin{array}{cc|cc}
{\bf (8,1)}_{0}^{++++} &
~{\bf (3,\overline{2})}_{-5/6}^{++--} &
~{\bf (\overline{3},1)}_{-2/3}^{+--+} & {\bf (3,2)}_{1/6}^{+-+-} \\
%\hline
{\bf (\overline{3},2)}_{5/6}^{++--} &
~{\bf (1,3)}_{0}^{++++} &
~{\bf (3,2)}_{1/6}^{+-+-} & {\bf (1,1)}_{1}^{+--+} \\
\hline
{\bf (3,1)}_{2/3}^{+--+} & ~{\bf (\overline{3},\overline{2})}_{-1/6}^{+-+-} &
~{\bf (8,1)}_{0}^{++++}
& {\bf (\overline{3},2)}_{5/6}^{++--} \\
%\hline
{\bf (\overline{3},\overline{2})}_{-1/6}^{+-+-} & ~{\bf (1,1)}_{-1}^{+--+} &
~{\bf (3,\overline{2})}_{-5/6}^{++--} &
{\bf (1,3)}_{0}^{++++}
\end{array}\right]_{10\times 10} ~,    
\end{eqnarray}
where the superscripts of the matrix elements indicate the eigenvalues 
of $P_1$, $P_2$, $P_3$, and $P_4$ respectively.
Here, to avoid too much clutter, we have omitted the two $U(1)$ generators 
(${\bf (1,1)_0^{++++}}$).   
As shown in Eqs.~(\ref{so10}) and (\ref{24}), they appear
in the diagonal part of the matrix (\ref{so10/z2z2}).

For future convenience, let us define the $SO(10)$ generator pieces   
appearing in Eq.~(\ref{so10/z2z2}) more succinctly,
\begin{eqnarray} \label{su5}
\left[\begin{array}{cc|cc}
{\bf G}^{++++} & ~{\bf Q'}^{++--} & 
~{\bf U^{c}}^{+--+} & {\bf Q}^{+-+-} \\
%\hline
~{\bf \overline{Q'}}^{++--} & ~{\bf W}^{++++} & 
{\bf Q}^{+-+-} & ~{\bf E^{c}}^{+--+} \\
\hline
~{\bf \overline{U}^{c}}^{+--+} & {\bf \overline{Q}}^{+-+-} &
{\bf G}^{++++} & ~{\bf \overline{Q'}}^{++--} \\
%\hline
{\bf \overline{Q}}^{+-+-} & ~{\bf \overline{E}^{c}}^{+--+} &
~{\bf Q'}^{++--} & ~{\bf W}^{++++}
\end{array}\right] ~, 
\end{eqnarray}
whose entries are in one to one correspondence 
to those of Eq.~(\ref{so10/z2z2}).  
Note that ${\bf Q}$ denotes ${\bf (3,2)}_{1/6}$, while ${\bf Q'}$ denotes  
${\bf (3,\overline{2})}_{-5/6}$.  
Similarly, the two $U(1)$ generators ${\bf (1,1)_0^{++++}}$, 
which were omitted in Eq.~(\ref{so10/z2z2}), are defined as 
\begin{eqnarray}
{\bf Y}^{++++}~~~ {\rm and}~~~ {\bf X}^{++++}~,  
\end{eqnarray}   
where ${\bf Y}$ corresponds to the hypercharge generator of SM.  
We identify the eigenvalues of the above generators with those of   
the associated gauge fields (and gauginos).    

Suppose we have an $S^1/(Z_2\times Z_2')$ orbifold compactification 
in 5D space-time.  
The two $Z_2$ elements among Eqs.~(\ref{p1})--(\ref{p4}) 
can be employed so as to embed the internal $Z_2\times Z_2'$ into       
the two presumed reflection symmetries for the extra space, 
$y\leftrightarrow -y$ and $y'\leftrightarrow -y'$ ($y'=y+y_c/2$).  
Two eigenvalues of $P_{i}$ could be interpreted as 
the parities (or boundary conditions) of the relevant fields 
under such reflections~\cite{hebecker}.   
Thus, the wave function of a field with parity $(+-)$, for instance,
must vanish on the brane at $y=y_c/2$ (B2),  
while it survives at $y=0$ brane (B1).  Only those fields assigned $(++)$
parities contain massless modes in their Kaluza-Klein (KK) spectrum.   
Thus, even though the bulk Lagrangian respects $SO(10)$, 
the effective low energy theory possesses a smaller gauge symmetry 
associated with the $(++)$ generators.     

If $P_1$ (identity) and one more $P_i$ ($i=2,3,4$) are taken  
as $Z_2\times Z_2'$ elements, 
the $SO(10)$ gauge symmetry breaks to 
$SU(5)\times U(1)_X$, $SU(5)'\times U(1)_X'$~\cite{5dflipped}, and 
$SU(4)_c\times SU(2)_L\times SU(2)_R$~\cite{5dps,ks,hdkim}, respectively.    
On the other hand, with two different $P_i$'s from among $\{P_2, P_3, P_4\}$,
$SO(10)$ can be broken to
$SU(3)_c\times SU(2)_L\times U(1)_Y\times U(1)_X$~\cite{ks2,hebecker2}, 
as illustrated in Figure 1.  
\begin{figure}
\begin{center}
\begin{picture}(225,225)(0,0)

%%%%%%%%%%%%%%%%%%%%%%%%%%%%%%%%%%%%%%%%%%%%%%%%%%%%%%%%%%%%%%%%%%%%%%%%%%

\CArc(112.5,131.25)(52.5,108,72)
\CArc(90,92.25)(52.5,182,167)
\CArc(135,92.25)(52.5,15,359)
\Text(112.5,161)[]{${\bf Q'}$, ${\bf \overline{Q'}}$}
\Text(62,78)[]{${\bf Q}$, ${\bf \overline{Q}}$}
\Text(163.5,82.5)[]{${\bf U^c}$, ${\bf \overline{U}^c}$}
\Text(163.5,71.25)[]{${\bf E^c}$, ${\bf \overline{E}^c}$}
\Text(112.5,114)[]{${\bf SM}$}
\Text(112.5,100)[]{${\bf \times U(1)_X}$}
\Text(112.5,183.75)[]{${\bf 5-1}$}
\Text(38.25,99)[]{${\bf 5'-1'}$}
\Text(187.5,99)[]{${\bf 4-2-2}$}
\Text(112.5,210)[]{${\bf SO(10)}$}
\Line(89,210)(7.5,210)
\Line(7.5,210)(7.5,15)
\Line(7.5,15)(220,15)
\Line(220,15)(220,210)
\Line(220,210)(135,210)

%%%%%%%%%%%%%%%%%%%%%%%%%%%%%%%%%%%%%%%%%%%%%%%%%%%%%%%%%%%%%%%%%%%%%%%%%%%

\end{picture}
\caption{A diagram showing the generators of $SO(10)$ and its subgroups 
schematically.    
$5-1$, $5'-1'$, $4-2-2$, and SM denote $SU(5)\times U(1)_X$, 
$SU(5)'\times U(1)_X'$, $SU(4)_c\times SU(2)_L\times SU(2)_R$, 
and the MSSM gauge group, respectively.}
\end{center}
\end{figure}
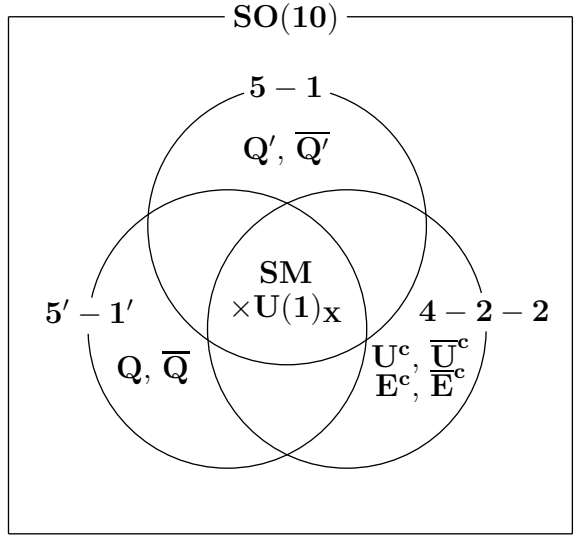

As is well known, compactification on $S^1/(Z_2\times Z_2')$ also 
can break the 4D $N=2$ SUSY to $N=1$.  
An $N=2$ supersymmetric vector multiplet is split into 
an $N=1$ vector multiplet and a chiral multiplet in adjoint representation.    
We assign the same parities as the generators 
to the associated vector multiplets as claimed above, 
but opposite parities to chiral multiplets.     
Then, $N=2$ SUSY is broken to $N=1$.   

\section{$SU(5)\times U(1)_X$~--~$SU(5)'\times U(1)_X'$}

Let us consider the case in which $P_2$ and $P_3$ operations 
are chosen as $Z_2$ and $Z_2'$ elements~\cite{ks2},  
corresponding to the second and the third parities  
in Eq.~(\ref{so10/z2z2}) and (\ref{su5}).     
With $P_2$, positive parities are assigned to the block-diagonal 
elements ($SU(5)\times U(1)_X$ generators and 
their associated gauge multiplets), 
while with $P_3$, positive parities are assigned to the generators of 
$SU(3)_c\times SU(2)_L\times U(1)_Y\times U(1)_X$ 
and ${\bf Q}^{-+}$, ${\bf \overline{Q}}^{-+}$, and 
to their associated gauge multiplets.  
Hence, after compactification, the gauge symmetry reduces 
to $SU(3)_c\times SU(2)_L\times U(1)_Y\times U(1)_X$.     
Together with ${\bf X}^{++}_V$, the $SU(5)$ gauge multiplets 
in Eq.~(\ref{su5}) survive at B1,  
\begin{eqnarray} \label{24}
{\bf 24}_V=\bigg({\bf G}_V^{++}+{\bf W}_V^{++}+{\bf Y}_V^{++}\bigg)
+\bigg({\bf Q'}_V^{+-}+{\bf \overline{Q'}}_V^{+-}\bigg)~~~ 
{\rm at~B1}~, 
\end{eqnarray} 
where the subscripts $V$ denote the vector multiplets. 
Thus $SU(5)\times U(1)_X$ should be preserved at B1~\cite{hebecker}.  

At B2 ${\bf Q'}_V^{+-}$ and ${\bf \overline{Q'}}_V^{+-}$ in Eq.~(\ref{24}) 
are replaced by ${\bf Q}_V^{-+}$ and ${\bf \overline{Q}}_V^{-+}$,  
which are in the ${\bf 10}_{-4}$ and ${\bf \overline{10}}_{4}$
representations of $SU(5)\times U(1)_X$,      
\begin{eqnarray}
{\bf 24}_V'=\bigg({\bf G}_V^{++}+{\bf W}_V^{++}+{\bf Y}_V^{++}\bigg)
+\bigg({\bf Q}_V^{-+}+{\bf \overline{Q}}_V^{-+}\bigg)~~~
{\rm at~B2}~.   
\end{eqnarray}
Note that the assigned hypercharges coincide 
with those given in `flipped' $SU(5)'\times U(1)_X'$~\cite{flipped}.  
The $U(1)_X'$ generator at B2, ${\bf X'}^{++}$ is defined as
\begin{eqnarray}
{\rm diag}({\bf 1}_{3\times 3}, -{\bf 1}_{2\times 2},
-{\bf 1}_{3\times 3}, {\bf 1}_{2\times 2}) ~.
\end{eqnarray}
Thus, the $U(1)_X'$ charges of the surviving elements at B2
turn out to be zero, while the other components are
assigned $-4$ or $4$.
The $U(1)_X'$ generator and the matrix elements with $(++)$, $(-+)$ parities
in Eq.~(\ref{so10/z2z2})
can be block-diagonalized to the form given in Eq.~(\ref{so10})
\begin{eqnarray} \label{flipped}
\left[\begin{array}{cc|cc}
{\bf G}^{++} & ~{\bf Q}^{-+} & 
~{\bf U^{c}}^{--} & {\bf Q'}^{+-} \\
%\hline
{\bf \overline{Q}}^{-+} & ~{\bf W}^{++} & 
{\bf Q'}^{+-} & ~{\bf \overline{E}^{c}}^{--} \\
\hline
~{\bf \overline{U}^{c}}^{--} & ~{\bf \overline{Q'}}^{+-} &
{\bf G}^{++} & ~{\bf \overline{Q}}^{-+} \\
%\hline
{\bf \overline{Q'}}^{+-} & ~{\bf E^{c}}^{--} &
{\bf Q}^{-+} & ~{\bf W}^{++}
\end{array}\right] ~,
\end{eqnarray}
through unitary transformation 
of the $SO(10)$ generator in Eq.~(\ref{so10/z2z2}) with 
\begin{eqnarray}
U_3=\left(\begin{array}{c|ccc}
I_{3\times 3} & 0 & 0 & 0 \\ \hline
0 & 0 & 0 & I_{2\times 2} \\
0 & 0 & I_{3\times 3} & 0 \\
0 & I_{2\times 2} & 0 & 0
\end{array}\right)_{10\times 10} ~.
\end{eqnarray}
In Eq.~(\ref{flipped}), the two superscripts denote the eigenvalues 
of $P_2$ and $P_3$.  
From Eq.~(\ref{flipped}), we conclude that 
the gauge symmetry at B2
is associated with a second (flipped) $SU(5)'\times U(1)_X'$
embedded in $SO(10)$~\cite{flipped}. 
%
%Indeed, if the $SU(5)$ in $SU(5)\times U(1)_X$
%respected at B1 is the Georgi-Glashow $SU(5)$,
%the $SU(5)'\times U(1)_X'$ preserved at B2 should be the `flipped' one
%$SU(5)'\times U(1)_X'$, in which the MSSM hypercharge
%is defined as $Y\equiv -(Y_{\rm GG}-Y_{X}')/5$ \cite{flipped}.
%Thus, in this paper we take $SU(5)\times U(1)_X$ at B1 and
%$SU(5)'\times U(1)_X'$ at B2. 

To break 4D $N=2$ SUSY, opposite parities should be assigned to 
the chiral multiplet $(\Phi+iA_5,\lambda_2)$, where $\Phi$, $A_5$, $\lambda_2$ 
belong to $N=2$ vector multiplets.  
The non-vanishing chiral multiplets at B1 are
\begin{eqnarray} \label{chiral1}
{\bf 10}_\Sigma &=& {\bf U^c}_\Sigma^{++}+{\bf E^c}_\Sigma^{++}
+{\bf Q}_\Sigma^{+-} ~,~  \\ \label{chiral1'}
{\bf \overline{10}}_\Sigma &=& {\bf \overline{U}^c}_\Sigma^{++}
+{\bf \overline{E}^c}_\Sigma^{++}
+{\bf \overline{Q}}_\Sigma^{+-} ~~~~ ~,
\end{eqnarray}
while on B2, ${\bf Q}_\Sigma^{+-}$ and ${\bf \overline{Q}}_\Sigma^{+-}$ 
are replaced by
${\bf Q'}_\Sigma^{+-}$ and ${\bf \overline{Q'}}_\Sigma^{+-}$   
(which are in ${\bf 24}_\Sigma$ and ${\bf 24'}_\Sigma$ at B1).  
Together with the vector-like pairs with $(++)$ parities, 
they compose ${\bf 10}_{-4}'$ and ${\bf \overline{10}_4}'$-plets
of $SU(5)'\times U(1)_X'$ at B2,  
\begin{eqnarray}
{\bf 10'}_\Sigma &=& {\bf U^c}_\Sigma^{++}+{\bf \overline{E}^c}_\Sigma^{++}
+{\bf Q'}_\Sigma^{-+} ~~,~  \\ 
{\bf \overline{10}'}_\Sigma &=& {\bf \overline{U}^c}_\Sigma^{++}
+{\bf E^c}_\Sigma^{++}
+{\bf \overline{Q'}}_\Sigma^{-+} ~. \label{chiral2} 
\end{eqnarray}
We note in Eqs.~(\ref{chiral1})--(\ref{chiral2}) the appearance of  
two vector-like pairs 
${\bf U^c}_\Sigma^{++}$, ${\bf \overline{U}^c}_\Sigma^{++}$ and 
${\bf E^c}_\Sigma^{++}$, ${\bf \overline{E}^c}_\Sigma^{++}$, 
which contain massless modes.   
We summarize the above results in Table I.  
\vskip 0.4cm
\begin{center}
\begin{tabular}{|c|c||cccccc|} \hline
Vector (B1) ~&~ ${\bf 24}_V$, ${\bf 1}_V$ ~&~  
${\bf G}_V^{++}$~, & ${\bf W}_V^{++}$~, & ${\bf Y}_V^{++}$~, &  
${\bf X}_V^{++}$~, & ${\bf Q'}_V^{+-}$~, & ${\bf Q'}_V^{+-}$~ 
\\
Chiral (B1) ~&~ ${\bf 10}_\Sigma$, ${\bf \overline{10}}_\Sigma$ ~&~  
${\bf U^c}_\Sigma^{++}$, & ${\bf E^c}_\Sigma^{++}$, & 
${\bf Q}_\Sigma^{+-}$~, & ${\bf \overline{U}^c}_\Sigma^{++}$, &  
${\bf \overline{E}^c}_\Sigma^{++}$, & ${\bf \overline{Q}}_\Sigma^{+-}$~ 
\\ \hline \hline
Vector (B2) & ${\bf 24'}_V$, ${\bf 1'}_V$ &~  
${\bf G}_V^{++}$~, & ${\bf W}_V^{++}$~, & ${\bf Y}_V^{++}$~, & 
${\bf X'}_V^{++}$, & ${\bf Q}^{-+}_V$~, & ${\bf \overline{Q}}_V^{-+}$~ 
\\
Chiral (B2) & ${\bf 10'}_\Sigma$, ${\bf \overline{10}'}_\Sigma$ &~  
${\bf U^c}_\Sigma^{++}$, & ${\bf \overline{E}^c}_\Sigma^{++}$, & 
${\bf Q'}_\Sigma^{-+}$~, & ${\bf \overline{U}^c}_\Sigma^{++}$, & 
${\bf E^c}_\Sigma^{++}$, & ${\bf \overline{Q'}}_\Sigma^{-+}$~
\\ \hline
\end{tabular}
\vskip 0.4cm
{\bf Table I.~} Surviving superfields on each brane 
in the $SO(10)$ gauge multiplet.
\end{center}
To preserve the successful MSSM gauge coupling unification,
we need to remove them from the low energy spectrum.  
To realize the MSSM gauge symmetry at lower energies, 
we employ the Higgs mechanism via bulk Higgs fields.  
This is because with brane Higgs fields, it is hard to provide heavy masses 
for the vector-like pairs,  
${\bf U^c}_\Sigma^{++}$, ${\bf \overline{U}^c}_\Sigma^{++}$, and 
${\bf E^c}_\Sigma^{++}$, ${\bf \overline{E}^c}_\Sigma^{++}$.     
Let us introduce two pairs of Higgs hypermultiplets 
${\bf 16}$, ${\bf \overline{16}}$ as shown in Table II.      
\vskip 0.4cm
\begin{center}
\begin{tabular}{|c||c|c|} \hline
Hypermultiplets & $Z_2\times Z_2'$~ parities & $U(1)_R$ 
\\ \hline \hline
${\bf 16}_H~$ & ~${\bf u^c}^{--},~{\bf e^c}^{--},~{\bf q}^{-+}~~;~~
{\bf d^c}^{++},~{\bf l}^{+-}~~;~~{\bf \nu^c}^{++}$ & $0$ 
\\
${\bf 16^c}_H$ & ${\bf u}^{++}~,~{\bf e}^{++}~,~{\bf q^c}^{+-}~;~~
{\bf d}^{--}~,~{\bf l^c}^{-+}~;~~{\bf \nu}^{--}$ & $0$ 
\\ \hline
${\bf \overline{16}}_H$ & ~${\bf \overline{u}^c}^{--},
~{\bf \overline{e}^c}^{--},~{\bf \overline{q}}^{-+}~~;~~
{\bf \overline{d}^c}^{++},~{\bf \overline{l}}^{+-}~~;~~
{\bf \overline{\nu}^c}^{++}$ &$0$ 
\\
${\bf \overline{16}^c}_H$ & ${\bf \overline{u}}^{++}~,
~{\bf \overline{e}}^{++}~,~{\bf \overline{q}^c}^{+-}~;~~
{\bf \overline{d}}^{--}~,~{\bf \overline{l}^c}^{-+}~;~~
{\bf \overline{\nu}}^{--}$ & $0$ 
\\ \hline 
\end{tabular}
\vskip 0.4cm
{\bf Table II.~} $Z_2\times Z_2'$ parities of the bulk Higgs hypermultiplets.
\end{center}
From ${\bf 16}_H$ and ${\bf \overline{16}}_H$, 
the surviving fields at B1 and B2 are 
\begin{eqnarray}
{\bf 16}_H&:&({\bf d^c}^{++},~ {\bf l}^{+-};~ {\bf \nu^c}^{++}) 
~~~{\rm at~~B1} ~, \\
 &&({\bf d^c}^{++},~{\bf q}^{-+},~{\bf \nu^c}^{++}) ~~~{\rm at~~B2} ~, \\
{\bf \overline{16}}_H&:&({\bf \overline{d}^c}^{++},~ {\bf \overline{l}}^{+-};~ 
{\bf \overline{\nu}^c}^{++}) ~~~{\rm at~~B1}~, \\
&&({\bf \overline{d}^c}^{++},{\bf \overline{q}}^{-+},
~{\bf \overline{\nu}^c}^{++})~~~~{\rm at~~B2} ~,
\end{eqnarray}
They compose (${\bf \overline{5}}_{-3}$; ${\bf 1}_{5}$) and 
(${\bf 5}_{3}$; ${\bf \overline{1}}_{-5}$) representations 
of $SU(5)\times U(1)_X$
at B1, and ${\bf 10}_{1}'$, ${\bf \overline{10}}_{-1}'$
of $SU(5)'\times U(1)_X'$ at B2.  

In order to realize $N=1$ SUSY, the surviving fields from  
${\bf 16^c}$, ${\bf \overline{16}^c}$ on the two branes should be 
as follows:    
\begin{eqnarray}
{\bf 16^c}_H&:&({\bf u}^{++},~ {\bf q^c}^{+-},~ {\bf e}^{++})
~~~{\rm at~~B1} ~, \\ 
&:& ({\bf u}^{++},~{\bf l^c}^{-+};~{\bf e}^{++}) ~~~{\rm at~~B2} ~, \\
{\bf \overline{16}^c}_H&:&({\bf \overline{u}}^{++},~{\bf \overline{q}^c}^{+-},~ 
{\bf \overline{e}}^{++})~~~{\rm at~~B1} ~, \\
&:& ({\bf \overline{u}}^{++},~{\bf \overline{l}^c}^{-+};
~{\bf \overline{e}}^{++}) ~~~{\rm at~~B2} ~.  
\end{eqnarray}
They compose ${\bf 10^c}_{-1}$, 
${\bf \overline{10}^c}_{1}$ ($={\bf 10}_{1}$) at B1,  and 
(${\bf \overline{5}^{c'}}_{3}$; ${\bf 1^{c'}}_{-5}$), 
(${\bf 5^{c'}}_{-3}$; ${\bf \overline{1}^c}_{5}$) at B2, 
respectively.  
The results are summarized in Table III.  
\vskip 0.4cm
\begin{center}
\begin{tabular}{|c|c|c|c|c|} \hline
B1 & ${\bf 5}_H$, ${\bf 1}_H$ & 
${\bf \overline{5}}_H$, ${\bf \overline{1}}_H$ & ${\bf 10^c}_H$ &   
${\bf \overline{10}^c}_H$ 
\\ \hline
 & ${\bf d^c}^{++}$, ${\bf l}^{+-}$, ${\bf \nu^c}^{++}$ &
${\bf \overline{d}^c}^{++}$, ${\bf \overline{l}}^{+-}$, 
${\bf \overline{\nu}^c}^{++}$ &
${\bf u}^{++}$, ${\bf e}^{++}$, ${\bf q^c}^{+-}$ &
${\bf \overline{u}}^{++}$, ${\bf \overline{e}}^{++}$, 
${\bf \overline{q}^c}^{+-}$ 
\\ \hline \hline
B2 & ${\bf 10'}_H$ & ${\bf \overline{10}'}_H$ &  
${\bf 5^{'c}}_H$, ${\bf 1^{'c}}_H$ &  
${\bf \overline{5}^{'c}}_H$, ${\bf \overline{1}^{'c}}_H$
\\ \hline
 & ${\bf d^c}^{++}$, ${\bf \nu^c}^{++}$, ${\bf q}^{-+}$ & 
${\bf \overline{d}^c}^{++}$, ${\bf \overline{\nu}^c}^{++}$, 
${\bf \overline{q}}^{-+}$ &  
${\bf u}^{++}$, ${\bf l^c}^{-+}$, ${\bf e}^{++}$  &  
${\bf \overline{u}}^{++}$, ${\bf \overline{l}^c}^{-+}$, 
${\bf \overline{e}}^{++}$  
\\ \hline
\end{tabular}
\vskip 0.4cm
{\bf Table III.~} Surviving Higgs superfields on the branes B1 and B2. 
\end{center}
   
Consider the following Higgs superpotentials on the two branes 
\begin{eqnarray} \label{sp1}
W_{B1}&=&\kappa_1S\bigg({\bf 16}_H{\bf \overline{16}}_H-M_1^2\bigg)~,  
\\ \label{sp2}
W_{B2}&=&\kappa_2S\bigg({\bf 16'}_H{\bf \overline{16}'}_H+{\bf 1}{\bf 1'}
-M_2^2\bigg) ~,  
\end{eqnarray}
where $\kappa_{1,2}$ ($M_{1,2}$) are dimensionless (dimensionful) parameters.  
Here ${\bf 16}_H{\bf \overline{16}}_H$    
stands for the superpotential couplings  
by the surviving Higgs at B1 shown in Table III,
${\bf 10^c}_{H}{\bf \overline{10}^c}_H+{\bf \overline{5}}_{H}{\bf 5}_{H}
+{\bf 1}_H{\bf \overline{1}}_{H}$ 
with arbitrary coefficients.  
${\bf 16'}_H{\bf \overline{16}'}_H$ in Eq.~(\ref{sp2})
is also similarly understood.
$S$ is a bulk singlet superfield with unit $U(1)_R$ charge, 
which can couple to the Higgs fields on both branes. 
Also, ${\bf 1}$, ${\bf 1'}$ are gauge singlet fields 
with suitable $U(1)_R$ charges.  
With non-zero vacuum expectation values (VEVs) of the scalar components 
of ${\bf \nu^c}^{++}$, ${\bf \overline{\nu}^c}^{c++}$, 
$SU(3)_c\times SU(2)_L\times U(1)_Y\times U(1)_X$ is spontaneously broken 
to the MSSM gauge group.  
Note that suitable VEVs of ${\bf 1}$, ${\bf 1'}$ can ensure
that the VEVs $\langle{\bf \nu^c}^{++}\rangle$ and
$\langle{\bf \overline{\nu}^c}^{c++}\rangle$ are constant 
along the extra dimension. 

With spontaneous symmetry breaking, the gauge bosons, gauge scalars and 
their superpartners in ${\bf 10}_{-4}$, ${\bf \overline{10}}_{4}$ 
acquire masses.         
The gauge bosons in ${\bf U^c}^{--}_V$, 
${\bf Q}^{-+}_V$, ${\bf E^c}^{--}_V$, and 
${\bf \overline{U}^c}^{--}_V$,                  
${\bf \overline{Q}}^{-+}_V$, ${\bf \overline{E}^c}^{--}_V$   
absorb a linear combination of $A_5$'s from   
\begin{eqnarray}
&&{\bf U^c}^{++}_\Sigma~ (n\neq 0)~,~~ 
{\bf Q}^{+-}_\Sigma~,~~{\bf E^c}^{++}_\Sigma~ (n\neq 0)~,~~~~{\rm and} 
\\ 
&&{\bf \overline{U}^c}^{++}_\Sigma~ (n\neq 0)~,~~                      
{\bf \overline{Q}}^{+-}_\Sigma~,~~{\bf \overline{E}^c}^{++}_\Sigma~ (n\neq 0)~, 
\end{eqnarray}
and from the Higgs fields    
\begin{eqnarray}
&&{\bf u^c}^{--}~,~~ {\bf q}^{-+}~,~~ {\bf e^c}^{--}~,~~~~{\rm and}
\\
&&{\bf \overline{u}^c}^{--}~,~~ {\bf \overline{q}}^{-+}~,
~~ {\bf \overline{e}^c}^{--}~. 
\end{eqnarray}  
The massless ($n=0$) modes of the gauge scalars $\Phi$, $A_5$s in     
${\bf U^c}^{++}_\Sigma$, ${\bf E^c}^{++}_\Sigma$, 
and ${\bf \overline{U}^c}^{++}_\Sigma$, ${\bf \overline{E}^c}^{++}_\Sigma$ 
obtain masses from the gauge coupling 
$g^2|\langle \nu_H^c\rangle A_5|^2$, where $\nu_H^c$ ($\nu_H^{c*}$) is 
the scalar component of ${\bf \nu^c}^{++}$, ${\bf \overline{\nu}^c}^{++}$.    
The gauge bosons in ${\bf Q'}_V^{+-}$ 
and ${\bf \overline{Q'}}_V^{+-}$ absorb the $A_5$'s from  
\begin{eqnarray}
{\bf Q'}_\Sigma^{-+}~, ~~ 
{\bf \overline{Q'}}_\Sigma^{-+}~. 
\end{eqnarray}  
We note that the gauge bosons absorb $A_5$'s 
carrying the same quantum numbers but opposite parities, 
whereas they absorb the Higgs fields with the same parities.  
This can be understood from the Lagrangian after symmetry breaking, 
${\cal L}\supset (\partial_5A_\mu-\partial_\mu A_5)^2\sim 
m_{KK}^2(A_\mu-\frac{1}{m_{KK}}\partial_\mu A_5)^2$ and 
${\cal L}\supset g^2v^2(A_\mu-\frac{1}{gv}\partial_\mu a)^2$, 
where $m_{KK}$ indicates the KK mass and $a$ is the Goldstone boson of 
the scalar Higgs $\phi=(v+\rho)e^{ia/v}/\sqrt{2}$.   

Finally, in order to realize the MSSM field contents at low energies, 
we should ensure that the three vector-like pairs of Higgs fields, 
${\bf u}^{++}$, ${\bf \overline{u}}^{++}$, 
${\bf d^c}^{++}$, ${\bf \overline{d}^c}^{++}$, and 
${\bf e}^{++}$, ${\bf \overline{e}}^{++}$ are heavy.  
This is possible, for example, by introducing at B1 
additional brane chiral superfields ${\bf 10}_{-1}^b$, 
${\bf \overline{10}}_1^b$, and ${\bf \overline{5}}_{-3}^b$, ${\bf 5}_3^b$ 
with unit $U(1)_R$ charges.       
(Gauge symmetry forbids their couplings to the chiral multiplets 
from the 4D $N=2$ vector multiplet.)  
%
%Only if their accompanied mass parameters are larger than $M_1$, 
%${\bf 16}_H$ and ${\bf \overline{16}}_H$ in Eq.~(\ref{sp1}) develop VEVs 
%in the neutrino direction.   
%

\section{$SU(5)\times U(1)_X$~--~$SU(4)_c\times SU(2)_L\times SU(2)_R$}

In this section, we take $Z_2\times Z_2'$ elements 
to be the $P_2$ and $P_4$.      
As already explained, by a $P_2$ operation,  
the $SU(5)\times U(1)_X$ generators are assigned positive 
parities and their associated gauge multiplets survive at B1.  
On the other hand, the $SO(10)$ generators with even parity under $P_4$ are  
\begin{eqnarray} \label{so6}
{\bf (8,1)}_{0}^{++}~,~~ {\bf (\overline{3},1)}_{-2/3}^{-+}~,~~ 
{\bf (3,1)}_{2/3}^{-+}~,~~{\bf (1,1)}_{0}^{++}~; \\
{\bf (1,3)}_{0}^{++}~,~~ {\bf (1,1)}_{1}^{-+}~,~~ {\bf (1,1)}_{-1}^{-+}~,~~
{\bf (1,1)}_{0}^{++}~,   \label{so4}
\end{eqnarray}
all of which survive at B2.  Here the superscripts denote $P_2$ and $P_4$ 
eigenvalues.  
The generators in Eqs.~(\ref{so6}) and (\ref{so4}) 
correspond to $SO(6)$ and $SO(4)$, respectively.  
To see this explicitly, we transform the $SO(10)$ generator 
in Eq.~(\ref{so10/z2z2}) with the unitary matrix, 
\begin{eqnarray}
U_4=\left(\begin{array}{c|cc|c}
I_{3\times 3} & 0 & 0 & 0 \\ \hline
0 & 0 & I_{2\times 2} & 0 \\
0 & I_{3\times 3} & 0 & 0 \\ \hline
0 & 0 & 0 & I_{2\times 2}
\end{array}\right)_{10\times 10} ~.  
\end{eqnarray}
The entries with even parities under $P_4$ are then block-diagonalized,  
\begin{eqnarray} \label{422generator}
\left[\begin{array}{cc|cc}
{\bf (8,1)}_{0}^{++} &
{\bf (\overline{3},1)}_{-2/3}^{-+} &
{\bf (3,\overline{2})}_{-5/6}^{+-} & {\bf (3,2)}_{1/6}^{--} \\
%\hline
{\bf (3,1)}_{2/3}^{-+} &
{\bf (8,1)}_{0}^{++} &
{\bf (\overline{3},\overline{2})}_{-1/6}^{--} &
{\bf (\overline{3},2)}_{5/6}^{+-} \\
\hline
{\bf (\overline{3},2)}_{5/6}^{+-} & {\bf (3,2)}_{1/6}^{--} &
{\bf (1,3)}_{0}^{++}
& {\bf (1,1)}_{1}^{-+} \\
%\hline
{\bf (\overline{3},\overline{2})}_{-1/6}^{--} &
{\bf (3,\overline{2})}_{-5/6}^{+-} &
{\bf (1,1)}_{-1}^{-+} &
{\bf (1,3)}_{0}^{++}
\end{array}\right]_{10\times 10} ~, 
\end{eqnarray}
where we have omitted the two $U(1)$ generators (${\bf (1,1)_0^{++}}$s)  
from the diagonal parts.  
Using Eq.~(\ref{general}), one can readily check that 
the two block-diagonal parts are 
$SO(6)\times SO(4)$ ($\sim SU(4)_c\times SU(2)_L\times SU(2)_R$) generators.  
The two off diagonal parts in Eq.~(\ref{422generator}) are identified 
with each other, and they compose the ${\bf (6,2,2)}$ representations 
under $SU(4)_L\times SU(2)_L\times SU(2)_R$.     
We conclude that by employing $P_2$ and $P_4$, $SU(5)\times U(1)_X$ and 
$SU(4)_L\times SU(2)_L\times SU(2)_R$ are preserved at B1 and B2, 
respectively.  The parities of $N=1$ gauge multiplets follow those of 
the corresponding generators. 
 
With opposite parities assigned to the chiral multiplets, 
the non-vanishing components at B1 are  
\begin{eqnarray}
{\bf 10}_\Sigma
&=&{\bf U^c}_\Sigma^{+-}
+{\bf E^c}_\Sigma^{+-} 
+{\bf Q}_\Sigma^{++}~,~~{\rm  and}  \\
{\bf \overline{10}}_\Sigma
&=&{\bf \overline{U}^c}_\Sigma^{+-}
+{\bf \overline{E}^c}_\Sigma^{+-}
+{\bf \overline{Q}}_\Sigma^{++} ~,   
\end{eqnarray}
while, on B2 brane, the surviving chiral multiplet is 
\begin{eqnarray}
{\bf (6,2,2)}_\Sigma 
={\bf Q}_\Sigma^{++}
+{\bf \overline{Q}}_\Sigma^{++}
+{\bf Q'}_\Sigma^{-+}
+{\bf \overline{Q'}}_\Sigma^{-+} ~.  
\end{eqnarray}     
Here we used the notations from Eq.~(\ref{su5}), and      
the subscript ``$\Sigma$'' stands for the chiral multiplet.  
We show in Table IV the surviving vector and chiral multiplets 
on each brane.   
%\vskip 0.4cm
\begin{center}
\begin{tabular}{|c|c||c|} \hline
Vector (B1) ~&~ ${\bf 24}_V$, ${\bf 1}_V$ ~&
${\bf G}_V^{++}$,~  ${\bf W}_V^{++}$,~  ${\bf Y}_V^{++}$;~ 
${\bf X}_V^{++}$,~  ${\bf Q'}_V^{+-}$,~  ${\bf Q'}_V^{+-}$
\\
Chiral (B1) ~&~ ${\bf 10}_\Sigma$, ${\bf \overline{10}}_\Sigma$ ~&
${\bf U^c}_\Sigma^{+-}$,~  ${\bf E^c}_\Sigma^{+-}$,~   
${\bf Q}_\Sigma^{++}$;~  ${\bf \overline{U}^c}_\Sigma^{+-}$,~ 
${\bf \overline{E}^c}_\Sigma^{+-}$,~  ${\bf \overline{Q}}_\Sigma^{++}$
\\ \hline \hline
Vector (B2) &${\bf 15}_V$, ${\bf 3}_V$, ${\bf 3'}_V$&
${\bf G}_V^{++}$,~${\bf U^c}^{-+}_V$,~${\bf \overline{U}^c}_V^{-+}$, 
${\bf X}_V^{++}$;~ ${\bf W}_V^{++}$;~  
${\bf E^c}^{-+}$,~${\bf \overline{E}^c}^{-+}$,~${\bf Y}_V^{++}$       
\\
Chiral (B2) ~&~ ${\bf (6,2,2)}_\Sigma$ ~&
${\bf Q}_\Sigma^{++}$,~  ${\bf \overline{Q}}_\Sigma^{++}$,~  
${\bf Q'}_\Sigma^{-+}$,~  ${\bf \overline{Q'}}_\Sigma^{-+}$   
\\ \hline
\end{tabular}
\vskip 0.4cm
{\bf Table IV.~} Surviving superfields on each brane 
in the $SO(10)$ gauge multiplet.
\end{center}

As seen from Table IV, the vector-like pair,  
${\bf Q}_\Sigma^{++}$ and ${\bf \overline{Q}}_\Sigma^{++}$ must be   
removed from the low energy spectrum.  
They can become massive through spontaneous symmetry breaking   
by the bulk Higgs.     
Table V shows the Higgs hypermultiplets and their quantum numbers. 
\vskip 0.4cm
\begin{center}
\begin{tabular}{|c||c|c|} \hline
Hypermultiplets & $Z_2\times Z_2'$~ parities & $U(1)_R$
\\ \hline \hline
${\bf 16}_H$ & ~${\bf u^c}^{-+},~{\bf e^c}^{-+},~{\bf q}^{--}~~;~~
{\bf d^c}^{++},~{\bf l}^{+-}~~;~~{\bf \nu^c}^{++} $ & $0$
\\
${\bf 16^c}_H$ & ${\bf u}^{+-}~,~{\bf e}^{+-}~,~{\bf q^c}^{++}~;~~
{\bf d}^{--}~,~{\bf l^c}^{-+}~;~~{\bf \nu}^{--}$ & $0$
\\ \hline
${\bf \overline{16}}_H$ & ~${\bf \overline{u}^c}^{-+},~
{\bf \overline{e}^c}^{-+},~{\bf \overline{q}}^{--}~~;~~
{\bf \overline{d}^c}^{++},~{\bf \overline{l}}^{+-}~~;~~
{\bf \overline{\nu}^c}^{++}$ & $0$
\\
${\bf \overline{16}^c}_H$ & ${\bf \overline{u}}^{+-}~,~
{\bf \overline{e}}^{+-}~,~{\bf \overline{q}^c}^{++}~;~~
{\bf \overline{d}}^{--}~,~
{\bf \overline{l}^c}^{-+}~;~~{\bf \overline{\nu}}^{--}$ & $0$
\\ \hline
\end{tabular}
\vskip 0.4cm
{\bf Table V.~} $Z_2\times Z_2'$ parities of the bulk Higgs hypermultiplets.  
\end{center}
Analogous to the previous case with $SU(5)\times U(1)_X-SU(5)'\times U(1)_X'$, 
at B1 they compose $SU(5)\times U(1)_X$ multiplets, 
${\bf 10}_{1}$, ${\bf \overline{5}}_{-3}$, ${\bf 1}_{5}$, etc.    
At B2 they compose $SU(4)_c\times SU(2)_L\times SU(2)_R$
multiplets such as ${\bf (4,2,1)}$ and ${\bf (\overline{4},1,2)}$ 
as shown in Table VI.  
\vskip 0.4cm
\begin{center}
\begin{tabular}{|c|c|c|c|} \hline
${\bf 5}_H$, ${\bf 1}_H$ (B1) &
${\bf \overline{5}}_H$, ${\bf \overline{1}}_H$ (B1) & ${\bf 10^c}_H$ (B1) &
${\bf \overline{10}^c}_H$ (B1)
\\ \hline
${\bf d^c}^{++}$, ${\bf l}^{+-}$, ${\bf \nu^c}^{++}$ &
${\bf \overline{d}^c}^{++}$, ${\bf \overline{l}}^{+-}$,
${\bf \overline{\nu}^c}^{++}$ &
${\bf u}^{+-}$, ${\bf e}^{+-}$, ${\bf q^c}^{++}$ &
${\bf \overline{u}}^{+-}$, ${\bf \overline{e}}^{+-}$, 
${\bf \overline{q}^c}^{++}$ 
\\ \hline \hline
${\bf (\overline{4},1,2)}_H$ (B2) & ${\bf (4,1,2)}_H$ (B2) &
${\bf (4^c,2,1)}_H$ (B2) & ${\bf (\overline{4}^c,2,1)}_H$ (B2)  
\\ \hline
${\bf u^c}^{-+}$, ${\bf e^c}^{-+}$, ${\bf d^c}^{++}$, ${\bf \nu^c}^{++}$&  
${\bf \overline{u}^c}^{-+}$, ${\bf \overline{e}^c}^{-+}$, 
${\bf \overline{d}^c}^{++}$, ${\bf \overline{\nu}^c}^{++}$& 
${\bf q^c}^{++}$, ${\bf l^c}^{-+}$ &
${\bf \overline{q}^c}^{++}$, ${\bf \overline{l}^c}^{-+}$ 
\\ \hline
\end{tabular}
\vskip 0.4cm
{\bf Table VI.~} Surviving Higgs superfields on the branes B1 and B2. 
\end{center}

On the two branes, the Higgs superpotentials are  
\begin{eqnarray}\label{sp3}
W_{B1}&=&\kappa_1S\bigg({\bf 16}_H{\bf \overline{16}}_H-M_1^2\bigg) ~, 
\\ \label{sp4}
W_{B2}&=&\kappa_2S\bigg({\bf 16^c}_H{\bf \overline{16}^c}_H+{\bf 1}{\bf 1'}
-M_2^2\bigg) ~, 
\end{eqnarray}
where we schematically wrote the vector-like couplings
of the Higgs multiplets on the two branes, 
${\bf 10^c}_H{\bf \overline{10}^c}_H
+{\bf \overline{5}}_H{\bf 5}_H+{\bf 1}_H{\bf \overline{1}}_H$
and ${\bf (\overline{4},1,2)}_H{\bf (4,1,2)}_H
+{\bf (4^c,2,1)}_H{\bf (\overline{4}^c,2,1)}_H$ with arbitrary coefficients
as ${\bf 16}_H{\bf \overline{16}}_H$ and
${\bf 16^c}_H{\bf \overline{16}^c}_H$, respectively.
The gauge singlet superfields ${\bf 1}$, ${\bf 1'}$ are introduced 
for the same reason as in section 3. 
As in the previous case, the VEVs of ${\bf \nu^c}^{++}$, 
${\bf \overline{\nu}^c}^{++}$ lead to the MSSM gauge symmetry, and 
generate mass terms of ${\bf X}_V^{++}$, ${\bf Q}_\Sigma^{++}$, and 
${\bf \overline{Q'}}_\Sigma^{++}$.    
Additional B1 brane superfields ${\bf 10}_1^b$, ${\bf \overline{10}}_{-1}^b$, 
and ${\bf \overline{5}}_{-3}^b$, ${\bf 5}_{3}^b$ with $U(1)_R$ charges of 
unity, and their bilinear couplings with the Higgs fields at B1 
could simply make ${\bf d^c}^{++}$, ${\bf \overline{d}^c}^{++}$,
${\bf q^c}^{++}$, ${\bf \overline{q}^c}^{++}$, etc. heavy.    
%
%heavier than $M_1$, 
%which stabilize the Higgs VEVs along the singlet neutrino direction.  
%

Finally another scenario one could consider is one with $SU(5)'\times U(1)_X'$ 
and $SU(4)_c\times SU(2)_L\times SU(2)_R$ at B1 and B2 respectively.  
We will not pursue this any further here.  
 
\section{Fermion Masses and Mixings}

Let us consider the $SU(5)\times U(1)_X-SU(4)_c\times SU(2)_L\times SU(2)_R$ 
model. 
We place the second and third generation quarks and leptons 
on $SU(5)\times U(1)_X$ brane (B1), and the first generation 
on $SU(4)_c\times SU(2)_L\times SU(2)_R$ brane (B2).  
%
%While the second and third generations of the quarks, leptons and 
%their superpartners are included in ${\bf 10}_1$, ${\bf \overline{5}}_{-3}$, 
%${\bf 1}_5$ representations of $SU(5)$, the first generation are in 
%${\bf (4,2,1)}$ ($\equiv {\bf L}$), ${\bf (\overline{4},1,2)}$ 
%($\equiv {\bf R}$) representations of $SU(4)_c\times SU(2)_L\times SU(2)_R$.  
%
We also introduce an $U(1)_F$ flavor symmetry and 
a singlet bulk flavon field `$F$' carrying $U(1)_F$ charge of $-1$,   
such that $\frac{\langle F\rangle}{M_*}=\epsilon\approx 0.2$, 
where $M_*$ denotes the fundamental scale.  
The $U(1)_F$ charge assignments of the MSSM fields are shown in Table VII. 
\vskip 0.4cm
\begin{center}
\begin{tabular}{|c|c||c|c|c|} \hline
Representation & Family & Fields & $U(1)_R$ & $U(1)_F$ 
\\ \hline \hline 
${\bf 10}_1$ & 3rd & ${\bf u^c}_3$, ${\bf e^c}_3$, ${\bf q}_3$ & $1/2$ & $0$ 
\\
(B1) & 2nd & ${\bf u^c}_2$, ${\bf e^c}_2$, ${\bf q}_2$ & $1/2$ & $2$
\\ \hline
${\bf \overline{5}}_{-3}$ & 3rd & ${\bf d^c}_3$, ${\bf l}_3$ & $1/2$ & $3$ 
\\
(B1) & 2nd & ${\bf d^c}_2$, ${\bf l}_2$ & $1/2$ & $3$ 
\\ \hline
${\bf 1}_{5}$ & 3rd & ${\bf\nu^c}_3$ & $1/2$ & $-$ 
\\
(B1) & 2nd & ${\bf\nu^c}_2$ & $1/2$ & $-$ 
\\ \hline \hline
${\bf (4,2,1)}$~~ (B2) & 1st & ${\bf q}_1$, ${\bf l}_1$ & $1/2$ & $3$
\\
${\bf (\overline{4},1,2)}$~~ (B2) & 1st &~ ${\bf u^c}_1$, ${\bf d^c}_1$, 
${\bf e^c}_1$, ${\bf \nu^c}_1$ ~& 
$1/2$ & $5$  
\\ \hline
${\bf (1,2,2)}_H$ (B2) & (Higgs) &~${\bf h_u}$, ${\bf h_d}$ ~& $0$ & $0$ 
\\ \hline
\end{tabular}
\vskip 0.4cm
{\bf Table VII.~} Quantum numbers of the MSSM matter introduced 
on the two branes.
\end{center}
Here we assigned all the left handed lepton doublets the same $U(1)_F$ charges 
to realize in our model the idea of the ``democratic approach''
to neutrinos~\cite{democratic,st}.   

We could introduce the MSSM Higgs fields in the bulk or on B2 
so as to avoid the notorious doublet-triplet problem.  
For simplicity, let us introduce them on B2 in the representation 
${\bf (1,2,2)}$ of $SU(4)_c\times SU(2)_L\times SU(2)_R$.   
The masses of the first generation quarks and leptons
are generated from the coupling ${\bf (4,2,1)(\overline{4},1,2)(1,2,2)}_H$.    
The mass terms of the second and third generations and 
the mixing terms between the first and the other two generations 
are possible by introducing heavy vector-like ${\bf 16}$ fields in the bulk 
and through interactions shown in Fig. 2~\cite{vectorfield}.   
Table VIII shows the $Z_2$ parities and other quantum numbers of the 
bulk ${\bf 16}$ fields. 
\vskip 0.4cm
\begin{center}
\begin{tabular}{|c||c|c|c|} \hline
Hypermultiplets & $Z_2\times Z_2'$~ parities & $U(1)_R$ & $U(1)_F$
\\ \hline \hline
${\bf 16_{I}}$ & ~$U^{c-+}_{I},~Q^{--}_{I}~,
~E^{c-+}_{I}~;~~D^{c++}_{I},~L^{+-}_{I}~~;~~N^{c+-}_{I}$ &
$1/2$ & $0$
\\
${\bf 16^c_{I}}$ & $U^{+-}_{I}~,~Q^{c++}_{I},~
E^{+-}_{I}~~;~~D^{--}_{I}~,~L^{c-+}_{I}~;~~N^{-+}_{I}$ &
$1/2$ & $0$
\\ \hline
${\bf \overline{16}_{I}}$ & ~$\overline{U}^{c-+}_{I},~
\overline{Q}^{--}_{I}~,~\overline{E}^{c-+}_{I}~;~~
\overline{D}^{c++}_{I},~\overline{L}^{+-}_{I}~~;~~
\overline{N}^{c+-}_{I}$ & $1/2$ & $0$
\\
${\bf \overline{16}^c_{I}}$ & $\overline{U}^{+-}_{I}~,
~\overline{Q}^{c++}_{I},~\overline{E}^{+-}_{I}~~;~~
\overline{D}^{--}_{I}~,~\overline{L}^{c-+}_{I}~;~~
\overline{N}^{-+}_{I}$ & $1/2$ & $0$
\\ \hline \hline
${\bf 16_{II}}$ & ~$U^{c--}_{II},~Q^{-+}_{II}~,~E^{c--}_{II}~;~~
D^{c+-}_{II},~L^{++}_{II}~~;~~N^{c++}_{II} $ & $1/2$ & $0$
\\
${\bf 16^c_{II}}$ & $U^{++}_{II}~,~Q^{c+-}_{II},~
E^{++}_{II}~~;~~D^{-+}_{II}~,~L^{c--}_{II}~;~~N^{--}_{II}$ & $1/2$ & $0$
\\ \hline
${\bf \overline{16}_{II}}$ & ~$\overline{U}^{c--}_{II},~
\overline{Q}^{-+}_{II}~,~\overline{E}^{c--}_{II}~;~~
\overline{D}^{c+-}_{II},~\overline{L}^{++}_{II}~~;~~\overline{N}^{c++}_{II}$ &
$1/2$ & $0$
\\
${\bf \overline{16}^c_{II}}$ & $\overline{U}^{++}_{II}~,
~\overline{Q}^{c+-}_{II},~\overline{E}^{++}_{II}~~;~~\overline{D}^{-+}_{II}~,~
\overline{L}^{c--}_{II}~;~~\overline{N}^{--}_{II}$ & $1/2$ & $0$
\\ \hline 
\end{tabular}
\vskip 0.4cm
{\bf Table VIII.~} Quantum numbers assigned to 
the vector-like hypermultiplets.
\end{center}
They compose on B1 the $SU(5)\times U(1)$ multiplets ${\bf 10}_1$, 
${\bf \overline{10}}_{-1}$, ${\bf \overline{5}}_{-3}$, ${\bf 5}_3$, 
${\bf 1}_5$, ${\bf 1}_{-5}$, etc., 
whereas on B2 they correspond to ${\bf (4,2,1)}$, ${\bf (\overline{4},2,1)}$, 
${\bf (\overline{4},1,2)}$, and ${\bf (4,1,2)}$.  
The $U(1)_R$ charge assignments allow for supersymmetric mass terms for them  
on the branes.      

The effective 4D Yukawa couplings between bulk and brane fields turn out to be 
given by $y\sqrt{\frac{M_c}{M_*}}$~\cite{volumesuppress}, 
where $y$ is a coefficient of order unity, 
and $M_c$ denotes the compactification scale.   
We assume that the effective 4D Yukawa couplings are all of order unity, 
and this will be justified in section 6. 
The resultant MSSM Yukawa couplings are
\vskip 0.6cm
\begin{eqnarray}
\begin{array}{ccc}
{\bf u^c}_1~{\bf u^c}_2~{\bf u^c}_3
\end{array}
~~~~~~~~~~~~~~
\begin{array}{ccc}
{\bf d^c}_1~{\bf d^c}_2~{\bf d^c}_3
\end{array}
~~~~~~~~~~~~~~~~
\begin{array}{ccc}
{\bf e^c}_1~{\bf e^c}_2~{\bf e^c}_3
\end{array}
~~~~~~~~
\nonumber \\
\begin{array}{c}
{\bf q}_1\\{\bf q}_2\\{\bf q}_3
\end{array}
\left(\begin{array}{ccc}
\epsilon^8 & \epsilon^5 & \epsilon^3 \\
\epsilon^7 & \epsilon^4 & \epsilon^2 \\
\epsilon^5 & \epsilon^2 & 1
\end{array}\right){\bf h_u}~,
\begin{array}{c}
{\bf q}_1\\{\bf q}_2\\{\bf q}_3
\end{array}
\left(\begin{array}{ccc}
\epsilon^5 & \epsilon^3 & \epsilon^3 \\
\epsilon^4 & \epsilon^2 & \epsilon^2 \\
\epsilon^2 & 1 & 1
\end{array}\right)\epsilon^{3}{\bf h_d}~,
\begin{array}{c}
{\bf l}_1\\{\bf l}_2\\{\bf l}_3
\end{array}
\left(\begin{array}{ccc}
\epsilon^5 & \epsilon^2 & 1 \\
\epsilon^5 & \epsilon^2 & 1 \\
\epsilon^5 & \epsilon^2 & 1
\end{array}\right)\epsilon^{3}{\bf h_d}~,
\label{yukawa}
\end{eqnarray}
where the mixing elements between the first and the last two generations
are obtained after decoupling of the heavy bulk ${\bf 16}$ fields. 
%
%Diagonalization of the above matrices yields~\cite{st}
%\begin{eqnarray} \label{b}
%m_t~:~m_c~:~m_u&\approx&1~:~\epsilon^4~:~\epsilon^8, ~ \\
%m_b~:~m_s~:~m_d&\approx&1~:~\epsilon^2~:~\epsilon^5, ~ \\
%m_\tau~:~m_\mu~:~m_e&\approx&1~:~\epsilon^4~:~\epsilon^8~, ~\\
%m_b~\sim ~m_\tau&\sim&\frac{\epsilon^{3}}{{\rm tan}\beta}~m_t ~,
%\end{eqnarray}
%where ${\rm tan}\beta\equiv\frac{\langle h_u\rangle}{\langle h_d\rangle}
%\sim O(1)$, and the CKM mixing angles turn out to be
%\begin{eqnarray} \label{f}
%V_{us}\sim \epsilon~,~~V_{cb}\sim \epsilon^2~,~~V_{ub}\sim \epsilon^3~.~
%\end{eqnarray}
%The results in Eqs.~(\ref{b})--(\ref{f}) are consistent with the observations.
%
\begin{figure}
\begin{center}
\begin{picture}(275,235)(25,10)

%%%%%%%%%%%%%%%%%%%%%%%%%%%%%%%%%%%%%%%%%%%%%%%%%%%%%%%%%%%%%%%%%%%%%%%%%%

\ArrowLine(50,220)(100,220)
\ArrowLine(150,220)(100,220)
\Line(150,220)(175,220)
\ArrowLine(50,170)(100,170)
\ArrowLine(150,170)(100,170)
\Line(150,170)(175,170)
\ArrowArcn(175,195)(25,90,0)
\ArrowArc(175,195)(25,270,360)
\DashArrowLine(250,195)(200,195){3}
\Line(103,223)(97,217)
\Line(97,223)(103,217)
\Line(153,223)(147,217)
\Line(147,223)(153,217)
\Line(103,173)(97,167)
\Line(97,173)(103,167)
\Line(153,173)(147,167)
\Line(147,173)(153,167)
\Text(75,232)[]{${\bf q}_{2,3}$}
\Text(125,232)[]{$Q^{c++}_I$}
\Text(175,232)[]{$\overline{Q}^{c++}_I$}
\Text(100,208)[]{$\epsilon^{2,0}M_*$}
\Text(150,208)[]{$M_*$}
\Text(75,182)[]{${\bf u^c}_{2,3}$}
\Text(125,182)[]{$U_{II}^{++}$}
\Text(175,182)[]{$\overline{U}_{II}^{++}$}
\Text(100,158)[]{$\epsilon^{2,0}M_*$}
\Text(150,158)[]{$M_*$}
\Text(225,207)[]{${\bf h_u}$}
\Text(150,135)[]{\bf (a)}
%%%%%%%%%%%%%%%%%%%%%%%%%%%%%%%%%%%%%%%%%%%%%%%%%%%%%%%%%%%%%%%%%%%%%%%%%%

\ArrowLine(50,75)(100,75)
\ArrowLine(150,75)(100,75)
\ArrowLine(150,75)(200,75)
\ArrowLine(250,75)(200,75)
\DashArrowLine(200,35)(200,75){3}
\Line(103,78)(97,72)
\Line(97,78)(103,72)
\Line(153,78)(147,72)
\Line(147,78)(153,72)
\Text(75,87)[]{${\bf e^c}_{2,3}$}
\Text(125,87)[]{$\overline{E}^{c++}_{II}$}
\Text(175,87)[]{$E^{c++}_{II}$}
\Text(225,87)[]{${\bf l}_1$}
\Text(217,55)[]{${\bf h_d}$}
\Text(100,63)[]{$\epsilon^{2,0}M_*$}
\Text(202,85)[]{$\epsilon^3$}
\Text(150,63)[]{$M_*$}
\Text(150,20)[]{$\bf (b)$}
%%%%%%%%%%%%%%%%%%%%%%%%%%%%%%%%%%%%%%%%%%%%%%%%%%%%%%%%%%%%%%%%%%%%%%%%%%
\end{picture}
\caption{The effective Yukawa couplings of the second and
third generation up-type quarks (a), the mixing terms
between the first and the other two generations of charged leptons (b).
The trilinear coupling with Higgs in (a) and (b) are present on B2.
The other elements of the MSSM fermion mass matrices are also generated
similarly.}
\end{center}
\end{figure}
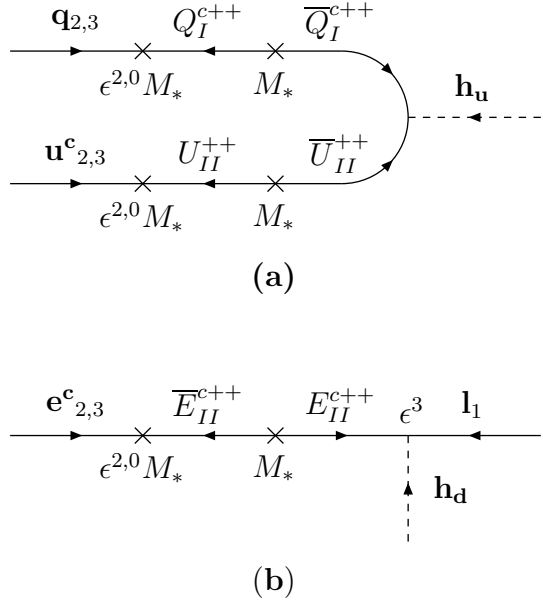
Diagonalization of the above matrices yields~\cite{st}
\begin{eqnarray} \label{b}
m_t~:~m_c~:~m_u&\approx&1~:~\epsilon^4~:~\epsilon^8, ~ \\
m_b~:~m_s~:~m_d&\approx&1~:~\epsilon^2~:~\epsilon^5, ~ \\
m_\tau~:~m_\mu~:~m_e&\approx&1~:~\epsilon^4~:~\epsilon^8~, ~\\
m_b~\sim ~m_\tau&\sim&\frac{\epsilon^{3}}{{\rm tan}\beta}~m_t ~,
\end{eqnarray}
where ${\rm tan}\beta\equiv\frac{\langle h_u\rangle}{\langle h_d\rangle}
\sim O(1)$, and the CKM mixing angles turn out to be
\begin{eqnarray} \label{f}
V_{us}\sim \epsilon~,~~V_{cb}\sim \epsilon^2~,~~V_{ub}\sim \epsilon^3~.~
\end{eqnarray}
The results in Eqs.~(\ref{b})--(\ref{f}) are consistent with the observations.

For the neutrino sector, as mentioned, we implement the democratic scenario 
presented in Ref.~\cite{democratic}.  
The contributions to the solar and atmospheric neutrino mixings 
from the charged lepton sector in Eq.(45) are of order unity~\cite{st}.   
The discussion in Ref.~\cite{st} shows how bilarge mixings are realized, 
taking into account the contributions from the Dirac and heavy Majorana 
sectors.  
To obtain the observed solar and atmospheric neutrino 
masses~\cite{solar,atmos},    
one could utilize the remaining undetermined $U(1)_F$ charges of 
the two right handed neutrino in Table VII.    
Note that the third mixing angle $\theta_{13}$ is expected 
in this approach to be not much smaller than 0.2 or so.  

%
%The expected lepton mixing angles are given by  
%\begin{eqnarray}
%{\rm sin}^22\theta_{\mu\tau}\sim {\rm sin}^22\theta_{e\mu}
%\sim {\rm sin}^22\theta_{e\tau} \sim 1~,    
%\end{eqnarray}
%which fit well with the SuperKamiokande experimental results 
%on atmospheric~\cite{atmos} and solar~\cite{solar} neutrino oscillations.   
%

Before closing this section, let us briefly discuss the $\mu$ term 
in the model.  
The $U(1)_R$ symmetry prevents a supersymmetric `bare' mass term of  
${\bf (1,2,2)}$ Higgs.    
Instead, we have the superpotential coupling  
\begin{eqnarray}
S{\bf (1,2,2)}_H{\bf (1,2,2)}_H ~.  
\end{eqnarray} 
The VEV $\langle S\rangle$ is zero for unbroken SUSY, 
but after SUSY breaking $\langle S\rangle$ becomes of order TeV, 
which induces the desired MSSM $\mu$ term~\cite{3221}.  

\section{Gauge Coupling Unification and Proton Decay}

In this section we discuss gauge coupling unification 
and proton stability in the model discussed in section 4 and 5.  
In higher dimensional SUSY GUT models, the compactification scale would be 
determined from the requirement that the MSSM gauge couplings should be 
unified at the cutoff scale $\Lambda$ ($\approx M_*$).    
Since the masses of the lightest colored gauge bosons 
such as $X$ and $Y$ gauge bosons in $SU(5)$ (${\bf Q'}$, ${\bf\overline{Q'}}$ 
in Fig. 1.) are in the compactification 
scale, in order to discuss the proton stability, 
we need to analyze the renormalization effects on the gauge couplings.   
But unfortunately in our model there are too many unknown parameters  
like $M_1$, $M_2$, and so on.       
%
%Moreover, the presence possibility of heavy vector-like field contents    
%on the $SU(4)_c\times SU(2)_L\times SU(2)_R$ brane 
%more obstruct the predictibility on the compactification scale.    
%

In our paper, to simplify the analysis, we make the assumption
that all mass parameters are very close to the cutoff scale $\Lambda$ 
including the spontaneous symmetry breaking scale 
on the branes~\cite{hdkim}.     
As discussed in section 3 and 4, 
the VEVs $\langle\nu^{c++}\rangle$, $\langle\overline{\nu}^{c++}\rangle$ 
could be constant along the extra dimension, and provide      
cutoff scale bulk masses to the vector and chiral multiplets 
charged under $U(1)_X$, belonging to 
${\bf 10}_{-4}$ and ${\bf\overline{10}}_4$ of $SU(5)$.  
We will keep only the vector and chiral multiplets of ${\bf 24}_0$ 
in the renormalization group (RG) analysis of the MSSM gauge couplings.   
Note that the first (second and third) generation Yukawa couplings 
still respect $SU(4)_c\times SU(2)_L\times SU(2)_R$ ($SU(5)\times U(1)_X$).   

In section 4 we introduced some vector-like pairs on B1 
to make ${\bf 5}_H$, ${\bf\overline{5}}_H$, ${\bf 10^c}_H$, and 
${\bf\overline{10}^c}_H$ in Table VI heavy (or heavier).   
The cutoff scale brane-localized bilinear couplings of $M_b$ 
shift their masses up by just about the compactification scale 
$M_c$ ($=\pi/y_c$)~\cite{localmass,hdkim}, 
so a lot of KK modes are not decoupled. 
The physical masses of the fields with $(++)$ $(+-)$ parities become   
\begin{eqnarray}
(++) ~&:& ~~2nM_c \longrightarrow (2n+1)(1-\delta)M_c ~, \\
(+-) ~&:& ~~(2n+1)M_c \longrightarrow (2n+2)(1-\delta)M_c ~,    
\end{eqnarray}
where $\delta$ $(<<1)$ is proportional to $M_c^2/M_b^2$.      
To simplify things, we set all brane-localized supersymmetric masses  
to be the same.   
Table IX shows the (shifted) mass spectrum for the bulk Higgs fields 
in Table V   
in the presence of brane localized superheavy mass terms.  
%\vskip 0.4cm
\begin{center}
\begin{tabular}{|c|c|c|c|c|} \hline
%%%%%%%%%%%%%
${\bf u^c}^{-+}$ & ${\bf e^c}^{-+}$ & ${\bf q}^{--}$ &
${\bf d^c}^{++}$ & ${\bf l}^{+-}$
\\ \hline
$2n+1$ & $2n+1$ & $2n+2$ &
$(2n+1)(1-\delta)$ & $(2n+2)(1-\delta)$
\\ \hline\hline
%%%%%%%%%%
${\bf u}^{+-}$ & ${\bf e}^{+-}$ & ${\bf q^c}^{++}$ &
${\bf d}^{--}$ & ${\bf l^c}^{-+}$
\\ \hline
$(2n+2)(1-\delta)$ & $(2n+2)(1-\delta)$ & $(2n+1)(1-\delta)$ &
$2n+2$ & $2n+1$
\\ \hline\hline
%%%%%%%%%%
${\bf \overline{u}^c}^{-+}$ & ${\bf \overline{e}^c}^{-+}$ &
${\bf \overline{q}}^{--}$ & ${\bf \overline{d}^c}^{++}$ &
${\bf \overline{l}}^{+-}$
\\ \hline
$2n+1$ & $2n+1$ & $2n+2$ &
$(2n+1)(1-\delta)$ & $(2n+2)(1-\delta)$
\\ \hline\hline
%%%%%%%%%%
${\bf \overline{u}}^{+-}$ & ${\bf \overline{e}}^{+-}$ &
${\bf \overline{q}^c}^{++}$ & ${\bf \overline{d}}^{--}$ &
${\bf \overline{l}^c}^{-+}$
\\ \hline
$(2n+2)(1-\delta)$ & $(2n+2)(1-\delta)$ & $(2n+1)(1-\delta)$ &
$2n+2$ & $2n+1$
\\ \hline
\end{tabular}
\vskip 0.4cm
{\bf Table IX.~} Mass spectrum of the bulk Higgs (normalized to  $M_c$).  
\end{center}
On the other hand, the corresponding brane fields 
${\bf \overline{5}}^b_{-3}$, ${\bf 5}^b_3$, ${\bf 10}^b_1$, and 
${\bf \overline{10}}^b_{-1}$ with unit $U(1)_R$ charges 
are simply decoupled due to the cutoff scale brane-localized 
supersymmetric mass terms.    

From the electroweak to the compactification scale, 
only massless modes of the bulk fields and light brane fields 
contribute to the RG equations of the MSSM gauge couplings.     
On the other hand, above the compactification scale, 
contributions from the KK modes of the bulk fields begin to appear,    
so that the MSSM gauge couplings show a linear dependence on 
the energy scale~\cite{hn03,hn11,kimslee}.    
Thus, above the compactification scale the evolutions are sensitive to 
the ultraviolet physics.  
However, the quantity  
$\Delta_i(\mu)\equiv\alpha_i^{-1}(\mu)-\alpha_1^{-1}(\mu)$ ($i=2,3$), 
displays logarithmic behavior, and can be meaningfully discussed 
even above the compactification scale~\cite{hn03,hn11}.  
In our model we have        
\begin{eqnarray} \label{RG1}
\Delta_3(m_Z)&=&\frac{1}{2\pi}\bigg[
%
%-\frac{3}{2}{\rm ln}\frac{m_{\rm SUSY}}{\mu}
-\frac{48}{5}~{\rm ln}\frac{\Lambda}{m_Z}
-6\sum_{n=0}^{N}{\rm ln}\frac{\Lambda}{(2n+2)M_c}
+6\sum_{n=0}^{N}{\rm ln}\frac{\Lambda}{(2n+1)M_c}\bigg]~, \\
\Delta_2(m_Z)&=&\frac{1}{2\pi}\bigg[
%
%-\frac{5}{3}{\rm ln}\frac{m_{\rm SUSY}}{\mu}
-\frac{28}{5}~{\rm ln}\frac{\Lambda}{m_Z}
-4\sum_{n=0}^{N}{\rm ln}\frac{\Lambda}{(2n+2)M_c}
+4\sum_{n=0}^{N}{\rm ln}\frac{\Lambda}{(2n+1)M_c}\bigg]~, \label{RG2}
\end{eqnarray}
with $\Lambda\gapproxeq (2N+2)M_c$.   
Here we set $\alpha_3=\alpha_2=\alpha_1$ at $\mu=\Lambda$.  
Note that the $\delta$ dependences in Eqs.~(\ref{RG1}) and (\ref{RG2})  
exactly cancel out.   
The beta function coefficients of the first terms 
in Eqs.~(\ref{RG1}) and (\ref{RG2}) are the same as 
in the MSSM case.  They result from contributions 
from the zero modes in ${\bf 24}_0$, and the brane matter fields 
of Table VII.   
Note that the vector-like superfields shown in Table VIII would draw
the MSSM gauge couplings much larger values upto the cutoff scale.
But they do not affect $\Delta_3(\mu)$ and $\Delta_2(\mu)$,
because the superfields with the same $Z_2\times Z_2'$ parities in Table VIII
compose complete ${\bf 16}$, ${\bf\overline{16}}$ of $SO(10)$.  

The linear combination $-[2\Delta_3(m_Z)+3\Delta_2(m_Z)]$ gives  
\begin{eqnarray} \label{linear1}
5\alpha_1^{-1}(m_Z)-3\alpha_2^{-1}(m_Z)-2\alpha_3^{-1}(m_Z) 
=\frac{1}{2\pi}\bigg[
%
%8~{\rm ln}\frac{m_{\rm SUSY}}{m_Z}+
36~{\rm ln}\frac{\Lambda}{m_Z}-24\sum_{n=0}^N\frac{2n+2}{2n+1}\bigg] ~,   
\end{eqnarray} 
which interestingly coincides with the result in Ref.~\cite{hn03}.     
Comparison with the corresponding linear combination 
in the usual 4D MSSM leads to 
\begin{eqnarray} 
{\rm ln}\frac{M_c}{m_Z}={\rm ln}\frac{M_U}{m_Z}
+\frac{2}{3}\sum_{n=0}^N{\rm ln}\frac{2n+2}{2n+1}-{\rm ln}(2N+2) ~, 
\end{eqnarray}
where $M_U$ lies in the range 
$1\times 10^{15}~{\rm GeV}\lapproxeq M_U\lapproxeq 3\times 10^{16}~{\rm GeV}$ 
from the experimental values of the gauge couplings.    
In section 5 we assumed that the effective Yukawa coupling  
$y\sqrt{\frac{M_c}{M_*}}$ ($\approx y\sqrt{\frac{M_c}{\Lambda}}$) is $O(1)$, 
where $y\sim O(1)$.   
If we take $\Lambda\approx 10M_c$ ($N=4$) as in Ref.~\cite{hn03}, 
which is also consistent with our assumption, 
$M_c$ is restricted by  
\begin{eqnarray} 
3\times 10^{15}~{\rm GeV}\lapproxeq M_c\lapproxeq 8\times 10^{15}~{\rm GeV}~.
\end{eqnarray} 
Hence, the compactification scale can be high enough 
to fulfill the bound $M_c\gapproxeq 5\times 10^{15}$ GeV 
arising from proton decay experiments and the constraints 
on the masses of the $X$, $Y$ gauge bosons 
in $SU(5)$~\cite{SK,hn03,newdim5}.  
Note that improvements by an order of magnitude of proton decay experiments 
would severely constrain our model or find proton decay! 
Finally note that baryon number violating dimension five operators are 
eliminated by $U(1)_R$~\cite{3221}.   

Another useful linear combination $7\Delta_3(m_Z)-12\Delta_2(m_Z)$ or 
\begin{eqnarray} \label{linear2}
5\alpha_1^{-1}(m_Z)-12\alpha_2^{-1}(m_Z)+7\alpha_3^{-1}(m_Z)
=\frac{1}{2\pi}\bigg[-6\sum_{n=0}^N{\rm ln}\frac{2n+2}{2n+1}\bigg]  
\end{eqnarray} 
provides a bound on $N$, but it is rather weak 
due to the experimental uncertainty of $\alpha_3(m_Z)$~\cite{hn03}.    
The experimental data prefers a positive value 
on the right hand side of Eq.~(\ref{linear2}).  
Let us discuss how this could arise from threshold corrections 
with an example.    
In contrast to 4D $SO(10)$, $SU(4)_c\times SU(2)_L\times SU(2)_R$ as well as 
$[SU(4)_c\times SU(2)_L\times SU(2)_R]/Z_2$ can be embedded in 5D $SO(10)$ 
compactified on $S^1/(Z_2\times Z_2')$. 
Hence, superfields such ${\bf (4,1,1)}$ 
($={\bf (3,1)}_{1/6}+{\bf (1,1)}_{-1/2}$) and ${\bf (\overline{4},1,1)}$ 
($={\bf (\overline{3},1)}_{-1/6}+{\bf (1,1)}_{1/2}$) 
carrying $U(1)_R$ charges of $1/2$ and their supersymmetric mass terms 
may be introduced on B2.  
While they leave intact Eq.~(\ref{linear1}), 
each pair yields an additional positive contribution to Eq.~(\ref{linear2}), 
\begin{eqnarray} \label{correction}
+\frac{1}{2\pi}\times \bigg[9~{\rm ln}\frac{\Lambda}{m_{\bf 4}}\bigg] ~,   
\end{eqnarray} 
where ${m_{\bf 4}}$ denotes the supersymmetric mass of ${\bf (4,1,1)}$ and  
${\bf (\overline{4},1,1)}$.\footnote{ 
The contributions from ${\bf (3,1)}_{1/6}$, ${\bf (1,1)}_{-1/2}$, 
${\bf (\overline{3},1)}_{-1/6}$, ${\bf (1,1)}_{1/2}$
to the evolutions of the three MSSM gauge couplings $\alpha_1^{-1}$, 
$\alpha_2^{-1}$ and $\alpha_3^{-1}$ are respectively given by  
$b_1=\frac{3}{5}[(\frac{1}{6})^2\times 3\times 2+(\frac{1}{2})^2\times 2]
=\frac{2}{5}$, $b_2=0$, $b_3=\frac{1}{2}[1\times 2]=1$ 
(upto a factor of $\frac{1}{2\pi}{\rm ln}\frac{\Lambda}{\mu}$). 
Hence, their contribution to $5\alpha_1^{-1}-3\alpha_2^{-1}-2\alpha_3^{-1}$ 
vanishes, while they yield Eq.~(\ref{correction}).   
}  
This can be employed to flip the sign of the right hand side of 
Eq.~(\ref{linear2}) without changing $M_c$ and $\Lambda$.  
Thus, by introducing two pairs of such fields with mass $\sim \Lambda/10$, 
$\alpha_3(m_Z)$ can be brought closer to the experimental value 
($0.117\pm 0.002$)~\cite{data} than the MSSM prediction.  
Alternatively, we can achieve the same result with two 
${\bf (6,1,1)}$'s    
($={\bf (3,1)}_{-1/3}+{\bf (\overline{3},1)}_{1/3})$ 
of $[SU(4)_c\times SU(2)_L\times SU(2)_R]/Z_2$  
with suitable $U(1)_R$ charges.\footnote{
The contributions from ${\bf (3,1)}_{-1/3}$, ${\bf (\overline{3},1)}_{1/3}$ 
to the evolution of $\alpha_1^{-1}$, $\alpha_2^{-2}$, $\alpha_3^{-1}$ are  
respectively $b_1=\frac{3}{5}[(\frac{1}{3})^2\times 3\times 2]=\frac{2}{5}$,   
$b_2=0$, $b_3=\frac{1}{2}[1\times 2]=1$ (upto a factor of 
$\frac{1}{2\pi}{\rm ln}\frac{\Lambda}{\mu}$).    
}  

We have confined our discussion of gauge coupling unification to the model 
with gauge groups $SU(5)\times U(1)_X-SU(4)_c\times SU(2)_L\times SU(2)_R$.  
We can expect that an analogous discussion on gauge coupling
unification for the $SU(5)\times U(1)_X-SU(5)'\times U(1)'_X$ model 
can be carried out by introducing heavy vector-like pair(s) of chiral fields.  
For instance, we can introduce ${\bf 10'}_1$ 
($={\bf (\overline{3},1)}_{1/3}+{\bf (3,2)}_{1/6}+{\bf (1,1)}_0$) and  
${\bf\overline{10}'}_{-1}$, or ${\bf 1'}_5$ ($={\bf (1,1)}_{1}$) 
and ${\bf\overline{1}'}_{-5}$  
on the $SU(5)'\times U(1)'_X$ brane.   
While the ${\bf 10'}_1$, ${\bf\overline{10}'}_{-1}$ pair shifts up 
$\alpha_2$ and $\alpha_3$ relative to $\alpha_1$, 
the ${\bf 1'}_5$, ${\bf\overline{1}'}_{-5}$ pair only contributes to $\alpha_1$ 
near the cut-off scale.~\footnote{
${\bf 10'}_1$ and ${\bf\overline{10}'}_{-1}$ give 
$b_1=\frac{3}{5}[(\frac{1}{3})^2\times 3\times 2
+(\frac{1}{6})^2\times 6\times 2]=\frac{3}{5}$, 
$b_2=\frac{1}{2}[3\times 2]=3$, $b_3=\frac{1}{2}[(1+2)\times 2]=3$.  
Hence, $b_1<b_2=b_3$.  
On the other hand, ${\bf 1'}_5$ and ${\bf\overline{1}'}_{-5}$ just give 
$b_1=\frac{3}{5}[1^2\times 2]=\frac{6}{5}$, $b_2=b_3=0$.  
}    
Thus, depending on details of the model, it seems that one can always  
achieve gauge coupling unification using such pairs. 
 
\section{Conclusion}

We have considered a variety of symmetry breakings obtained from 
compactifying 5D $SO(10)$ on $S^1/(Z_2\times Z_2')$.  
In particular, the residual symmetry after compactification is 
$SU(3)_c\times SU(2)_L\times U(1)_Y\times U(1)_X$.  
We have seen how the MSSM can be realized at low energies  
after spontaneous breaking of $U(1)_X$.  
We have presented the implication of one particular example, 
in which the Higgs breaking scale of $U(1)_X$ is comparable to the cutoff 
scale.  Thus, effectively we have the breaking 
$SO(10)\rightarrow SU(5)\rightarrow$ MSSM, for which we study 
the implications for fermion masses and mixings, 
gauge coupling unification and proton decay.

\vskip 0.3cm
\noindent {\bf Acknowledgments}

\noindent
The work is partially supported
by DOE under contract number DE-FG02-91ER40626.

\end{document}